\documentclass[twocolumn,aps,prl,superscriptaddress,showpacs]{revtex4}

\usepackage{graphicx}% include figure files
\usepackage[verbose=true]{epstopdf}
\usepackage{hyphenat} % allows to break compound words between lines use: \hyp{}

\begin{document}

% Definitions
\newcommand{\etal}{\emph{et al.}}
\newcommand{\Li}{{}^{6}\textrm{Li}}
\newcommand{\LiTwoSthreetwo}{{2S_{1/2}}}
\newcommand{\UVtransition}{{2S_{1/2}\rightarrow 3P_{3/2}}}
\newcommand{\REDtransition}{{2S_{1/2}\rightarrow 2P_{3/2}}}
\newcommand{\GammaTwoP}{\Gamma_{2\mathrm{P}}}
\newcommand{\GammaThreeP}{\Gamma_{3\mathrm{P}}}
\newcommand{\Kfourty}{{}^{40}\textrm{K}}
\newcommand{\Kthirtynine}{{}^{39}\textrm{K}}
\newcommand{\kB}{k_{\mathrm{B}}}
\newcommand{\Limp}{\textrm{Li}\left|3/2,3/2\right\rangle}
\newcommand{\LiOneOne}{\textrm{Li}\left|1/2,1/2\right\rangle}
\newcommand{\IsatTwoP}{I_{\mathrm{sat}}^{2\mathrm{P}}}
\newcommand{\IsatThreeP}{I_{\mathrm{sat}}^{3\mathrm{P}}} % use \text{} instead of \textrm{}
\newcommand{\UVMOT}{\textrm{UV\,MOT}} % use \text{} instead of \textrm{}

\title{All\hyp{}optical production and transport of a large ${}^{6}\textrm{Li}$ quantum gas in a crossed optical dipole trap}

\author{Ch.~Gross}
\affiliation{Centre for Quantum Technologies (CQT), 3 Science Drive 2, Singapore 117543}
\author{H.~C.~J.~Gan}
\affiliation{Centre for Quantum Technologies (CQT), 3 Science Drive 2, Singapore 117543}
\author{K.~Dieckmann}
\email[Electronic address:]{phydk@nus.edu.sg}
\affiliation{Centre for Quantum Technologies (CQT), 3 Science Drive 2, Singapore 117543}
\affiliation{Department of Physics, National University of Singapore, 2 Science Drive 3, Singapore 117542}

\date{\today}

\begin{abstract}
We report on an efficient production scheme for a large quantum degenerate sample of fermionic lithium. The approach is based on our previous work on narrow-line $\UVtransition $ laser cooling resulting in a high phase-space density of up to $3\times10^{-4}$. This allows utilizing a large volume crossed optical dipole trap with a total power of $45\,\textrm{W}$, leading to high loading efficiency and $8\times10^6$ trapped atoms. The same optical trapping configuration is used for rapid adiabatic transport over a distance of $25\,\textrm{cm}$ in $0.9\,\textrm{s}$, and subsequent evaporative cooling. With optimized evaporation we achieve a degenerate Fermi gas with $1.7\times 10^{6}$ atoms at a temperature of $60 \, \textrm{nK}$, corresponding to $T/T_{\text{F}}=0.16\left(2 \right)$. Furthermore, the performance is demonstrated by evaporation near a broad Feshbach resonance creating a molecular Bose\hyp{}Einstein condensate of $3\times10^5$ lithium dimers.
\end{abstract}

% (Pacs)
% PACS 2010
% 37.10.De Atom cooling methods
% 37.10.Gh Atom traps and guides
% 67.85.Lm, degenerate Fermi gases
%67.85.Hj 	Bose-Einstein condensates in optical potentials

\maketitle

%%% Introduction -------------------------------------------------------------------------------

\section{Introduction}

%general
Experiments investigating quantum degenerate atomic gases have seen tremendous progress over the past two decades. As many new and complex experimental techniques for scientific exploration have been implemented, there is continued interest in the simplification of methods for the production of quantum degenerate samples. 

% Fermionic quantum gases based on improved cooling
In particular, quantum gases have been widely studied with lithium and potassium as they both have fermionic and bosonic isotopes. For these two species there is the additional difficulty that standard sub\hyp{}Doppler laser cooling is absent or inefficient. However, more recently alternative types of laser cooling like narrow\hyp{}line cooling \cite{Duarte2011b, McKay2011a} and gray optical molasses techniques \cite{Fernandes2012a,Grier2013,Salomon2013,Burchianti2014,Salomon2014} were successfully demonstrated. This lead to significantly improved phase\hyp{}space densities after laser cooling and the improved all\hyp{}optical production of quantum degenerate samples. Especially, the loading of the atoms into optical dipole traps can be improved, as this typically suffers from insufficient matching of trapping volumes and depths. For the case that efficient laser cooling cannot be attained, methods like high-power laser based traps \cite{OHara1999, Barrett2001} or resonator enhanced dipole traps \cite{Mosk2001} for larger trap depth and volume were previously developed. 

% access and transport
Another important aspect is the transport of the atomic clouds over long distances facilitating advanced techniques in manipulating and probing ultracold gases that require good optical or mechanical access. This mitigates constraints generally imposed by magneto\hyp{}optical trapping and pre\hyp{}cooling. Several transport methods were developed, which include the translation of magnetic traps \cite{Lewandowski2003, Greiner2001} and optical tweezers \cite{Gustavson2001, Zimmermann2011, Couvert2008, Leonard2014}, as well as $1\textrm{D}$ moving optical lattices \cite{Schmid2006, Middelmann2012}. In most of these experimental schemes the confinement of the atoms is optimized for the transport itself, which requires different measures for the loading of the atoms into the optical dipole trap and subsequent optimized evaporative cooling \cite{OHara2001, Luo2006}.

%%% Here we present... ----------------
Here we present an apparatus for the efficient production of a large quantum degenerate sample of fermionic lithium. Our experimental approach is based on our previous work on efficient narrow-line laser cooling and trapping on a transition in the ultraviolet (UV) regime \cite{Sebastian2014}. Due to the small linewidth, temperatures are achieved that lie well below the Doppler limit for standard cooling on the D2 transition resulting in high phase-space density samples In addition, the cooling remains effective when the sample is loaded into an optical dipole trap, and hence high loading efficiencies are obtained.

We employ a small\hyp{}angle crossed beam trapping geometry with a large volume, leading to good loading efficiencies from the ultraviolet magneto-optical trap (UV MOT), while simultaneously offering a suitable axial confinement for a fast transport of the atomic cloud. With this simple extension of the single beam optical tweezers method no further optical trapping steps are needed, such as a resonator enhanced ODT \cite{Zimmermann2011} or a transfer between subsequent ODTs of decreasing volume \cite{Omran2015}. Our trapping configuration is used to transport the atoms into a science cell in order to facilitate good optical access. Furthermore, the optical confinement is sufficient for efficient evaporation of the trapped atoms into the quantum degenerate regime. To underline the overall performance of this approach we present the production of a large Bose\hyp{}Einstein condensate (BEC) of Feshbach molecules.

%%% Experimental Approach ------------------------------------------------------------------------------

\section{Experimental Approach}

\begin{figure}
\includegraphics[width=8.0cm]{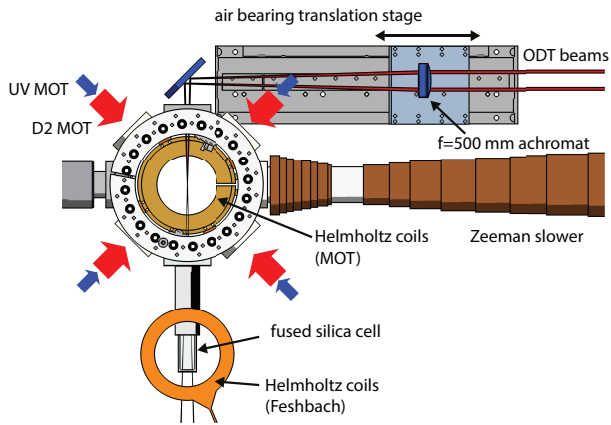}%
	\caption{(Color online) Schematic overview of our $\Li$ apparatus. The atoms are loaded from a Zeeman\hyp{}slowed atomic beam into a MOT on the D2 transition. The atomic cloud is further cooled and compressed on the narrow UV transition and then transferred into a crossed ODT comprising of two laser beams focused by a single lens. Subsequently, the atomic cloud is transported from the MOT chamber into an attached fused silica cell by moving the lens with an air bearing stage. A magnetic bias field is applied to access a Feshbach resonance for efficient evaporation of the atomic cloud into the quantum degenerate regime.}
	\label{fig:LithiumSetup}
\end{figure}

%%% UV MOT
Our experimental sequence for the production of a quantum degenerate $\Li$ gas starts with a two\hyp{}stage MOT, which was described in \cite{Sebastian2014}. At first, we employ a standard MOT on the $\textrm{D}2$ transition to obtain a large number of atoms. This is followed by Doppler cooling on the narrow UV $\UVtransition$ transition at $323\, \textrm{nm}$ with a natural linewidth of the excited state of $ \GammaThreeP = \left( 2 \pi \right) \times 754\, \textrm{kHz}$. After this two\hyp{}stage MOT, we typically obtain a $\Li$ cloud containing around $1.5 \times 10^{8}$ atoms at a temperature of $\sim50\,\mu$K with a central density on the order of $3\,\times 10^{11} \textrm{cm}^{-3}$.

%%% loading
In the next step the atoms are transferred into a crossed optical dipole trap that comprises of two beams intersecting at a small angle. The laser wavelength of $1070\, \textrm{nm} $ for the optical dipole trap is near a magic wavelength, where the differential ac Stark shift of the UV MOT transition vanishes \cite{Safronova2012, Duarte2011b}. This permits continued cooling on the UV transition during the initial stage of optical trapping and allows to take full advantage of the high phase\hyp{}space density of the UV MOT.

%%% transport
After loading we transport the atoms into a small glass cell attached to the MOT vacuum chamber as shown in Fig.~\ref{fig:LithiumSetup}. As the atoms are loaded from a Zeeman\hyp{}slowed atomic beam, the vacuum background is low and no differential pumping stage between the MOT and science cell is required. This allows the use of a linear optical transport. Our particular scheme demonstrates the displacement of a crossed optical dipole trap across a long distance, realized with an air bearing translation stage. This is a key feature of our simplified experimental approach, as efficient direct loading from the MOT, loss\hyp{}free near\hyp{}adiabatic transport, and efficient evaporative cooling are achieved in the same trap. 

%%% degeneracy 
Evaporative cooling of the atomic cloud is performed in the science cell with a spin mixture of the two energetically lowest Zeeman levels $\left|\textrm{F}=1/2,m_{F}=\pm 1/2\right\rangle$, commonly referred to as states $ \left| 1 \right\rangle$ and $ \left| 2 \right\rangle$. In order to tune the s\hyp{}wave scattering length $a_{12} $, magnetic bias fields of up to $\sim 1300\, \textrm{G}$ can be applied. For this work evaporative cooling is either carried out near the broad Feshbach resonance at $832 \, \textrm{G}$ \cite{Zurn2013} to create a strongly interacting Fermi quantum gas in the BEC\hyp{}BCS crossover regime \cite{Bartenstein2004}, or at around $330 \, \textrm{G}$ to obtain a weakly interacting Fermi gas.

%%% Loading of the ODT from the UV MOT----------------------------------------------------------------------
\section{Loading of the ODT from the UV MOT}

%%% trap description -----------
The power for ODT is derived from a single mode $1070\, \textrm{nm} $ fiber laser (IPG, YLR\hyp{}50\hyp{}1070\hyp{}LP) operated at a fixed nominal output power of $ 50 \, \textrm{W} $, that is equally split into two beams for the crossed ODT. For each beam the power is regulated by an acousto\hyp{}optic modulator (AOM). The crossed ODT is formed by focusing the two collimated parallel beams with $1/e^2$ waist of $2.6\, \textrm{mm} $ using a single $f=500\, \textrm{mm} $ lens, as shown in Fig.~\ref{fig:LithiumSetup}. The beams are symmetrically displaced by $ 13\, \textrm{mm} $ from the center of the lens, leading to a crossing angle of $3^{\circ}$ degrees. This angle is determined under the geometric constraint of the aperture of the vacuum viewport. The two beams intersect near their foci, which have a waist of $ 66\,\mu\textrm{m} $. We observed a slight power dependent astigmatism. In contrast to a single beam dipole trap, this has a negligible effect on the position of the ODT during forced evaporation, as the axial position of the trap is predominantly defined by the crossing of the two beams. 

The full trap depth $ U_{0} $ of the crossed trap with $ 22.5\, \textrm{W} $ per beam is $ k_{B}\times 330 \, \textrm{$\mu$K} $, where $ k_{{B}} $ is the Boltzmann constant, and the axial and radial trap frequencies are $ \omega_{z} = \left( 2 \pi \right) \times 81 \, \textrm{Hz} $ and $ \bar{\omega}_{r} = \sqrt{ \omega_x \omega_y }  = \left( 2 \pi \right) \times 3.04 \, \textrm{kHz} $, respectively. The trap frequencies were experimentally determined by observing the center of mass oscillations of the atomic cloud after a displacement from the trap center, induced by a magnetic field gradient pulse. From a series of such measurements we obtained the scaling of the trap frequencies with the power per beam $ P $ of the ODT, given by $ \omega_{z} = \left( 2\pi \right) \times 17.1 \, \sqrt{P} \,\textrm{Hz} $ and $ \bar{\omega}_{r} = \left( 2\pi \right) \times 641 \, \sqrt{P}\,\textrm{Hz} $. Hence, the effect of the power dependence of the astigmatism on the trap frequencies were found to be negligible. All relevant parameters of the ODT are summarized in Table~\ref{tab:TransportTable}.

\begin{table}[htbp]
	\centering
	\caption{Parameters of the implemented crossed ODT are compared to calculated values for a single beam configuration with identical total optical power and axial confinement. As reference the respective values are listed for an optical trap geometry used for transport as reported in Ref.~\cite{Zimmermann2011}. The values given for our crossed trap include small effects of imperfect alignment.}
		\begin{tabular}{l c  c  c}
		\hline \hline
		 & \hspace{0.125cm} crossed \hspace{0.125cm}   &   \hspace{0.125cm} single beam \hspace{0.125cm}    & \hspace{0.125cm} single beam  \hspace{0.25cm} \\ 
		 & ODT & ODT & ODT \cite{Zimmermann2011} \\
		\hline 
		$1/e^2$ waist & $ 66 \, \textrm{$\mu$m} $  & $ 36 \, \textrm{$\mu$m} $ & $ 21 \, \textrm{$\mu$m} $\\
		optical power & $ 45\, \textrm{W} $  & $ 45\, \textrm{W} $ & $ 2\, \textrm{W} $ \\ 
		$ U_{0}  / k_{B} $ & $ 330 \, \textrm{$\mu$K} $  & $ 1.3 \, \textrm{mK} $ & $ 173 \, \textrm{$\mu$K} $\\
		$\Gamma_{sc}$ & $ 0.46\, \textrm{Hz} $ & $ 1.9\, \textrm{Hz} $ & $ 0.24\, \textrm{Hz} $ \\ 
		$\omega_{z} / 2 \pi $ & $ 81\, \textrm{Hz} $  & $ 81\, \textrm{Hz} $ & $ 85\, \textrm{Hz} $ \\ 
		$\bar{\omega}_{r} / 2 \pi $ & $ 3.04\, \textrm{kHz} $ & $ 12.10\, \textrm{kHz} $ & $ 7.43\, \textrm{kHz} $ \\
		aspect ratio & $ 37.5 $ & $ 149 $ & $ 87.2 $ \\ 
		\hline \hline
		\end{tabular}
	\label{tab:TransportTable}
\end{table}

%%% Loading of ODT --------------
For the transfer of the atoms into the crossed ODT, the optical power is linearly ramped up within $ 2\, \textrm{ms}$ to the full trap depth during the compressed UV MOT phase. As described in \cite{Sebastian2014}, the peak density reaches a transient maximum value during this stage, at which point the ramp-up is initiated. After a combined loading and cooling duration of $ 10 \, \textrm{ms} $, the UV MOT is switched off by extinguishing the repumping light $ 0.5 \, \textrm{ms} $ in advance of the cooling light. This optically pumps the atoms into the $ F=1/2 $ hyperfine ground state manifold, creating an incoherent balanced mixture of the $ \left| 1 \right\rangle$ and $ \left| 2 \right\rangle$ states. For this work the UV MOT is realized with the repumping light tuned to the $\textrm{D}2$  transition for reasons of stability in switching the optical power for hyperfine pumping. Employing the UV repumping light instead, leads to a very similar loading efficiency.

%%% after loading we find, plain evaporation  ---------------
Immediately after the loading we find an atom number of $ N = 1.1\,\times 10^{7}$, which corresponds to a transfer efficiency of approximately $7\,\%$. The temperature is observed to be anisotropic with a geometric mean of $ T = 43 \, \textrm{$\mu$K} $. This results in an initial ratio of $\eta=7.7$ of the trap depth to the temperature of the atomic cloud, referred to as the truncation parameter. Following the loading of the atoms into the ODT we allow for $1 \,\textrm{s} $ of plain evaporation. For the purpose of thermalization a magnetic bias field of approximately $ 200\,\textrm{G}$ is applied, generated by inverting the polarity of one of the MOT coils. This tunes the s\hyp{}wave scattering length to $ a_{12} = -220 \, a_{0} $ between the two states \cite{Zurn2013}, where $ a_{0} $ is Bohr's radius. After plain evaporation the ODT typically contains a total number of atoms of $ N = 8\,\times 10^{6}$ at a temperature of $T = 30 \, \textrm{$\mu$K} $ corresponding to $T/T_F=2.4$. The Fermi temperature is defined as $k_B T_F = \hbar \bar{ \omega }  \left( 3N \right)^{1/3} $, where $ \bar{ \omega } = \left( \bar{\omega}_r^2 \, \omega_{z} \right) ^{1/3} \, $ is the mean trap frequency and $ \hbar $ is the reduced Planck constant. We estimate a systematic uncertainty of $ 10\, \% $ in all measured quantities.

\begin{figure}
\includegraphics[width=8.0cm]{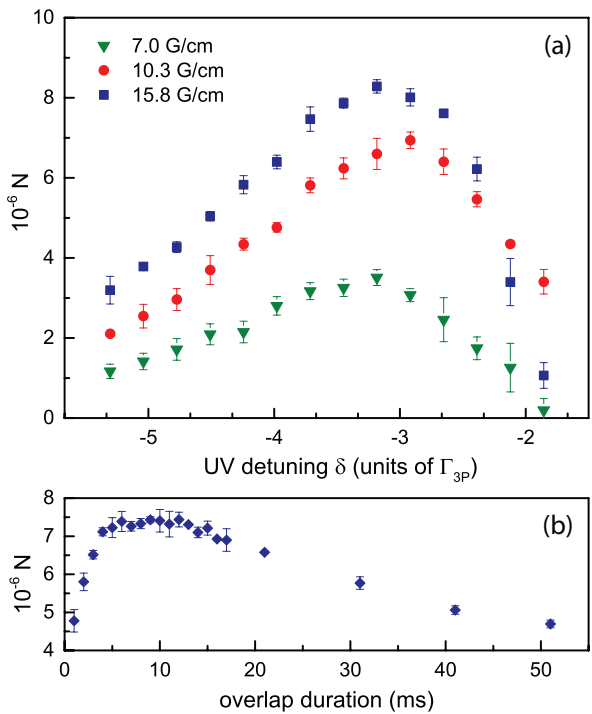}%
\caption{(Color online) (a) Total number of atoms $N$ in the ODT after plain evaporation as a function of the laser detuning with respect to the UV cooling transition for different axial magnetic field gradients of the MOT. The highest atom number is achieved with a detuning of $\delta\simeq\mathrm{-3.2} \, \Gamma_{\mathrm{3P}}$ at a gradient of $ 15.8 \, \textrm{G} / \textrm{cm} $. An increase of the magnetic field gradient beyond this value does not further improve the atom number. (b) Atom number vs. duration of continued UV cooling after the ODT is ramped up to full power. The initial increase in atom number demonstrates the effectiveness of UV cooling in presence of the $1070 \, \textrm{nm} $ trapping light, while the decrease at longer overlap durations is attributed to density dependent losses (see text). The error bars account for the statistical uncertainty.}
\label{fig:NvsDetuning}
\end{figure}

%%% atom number vs detuning   --------------
We investigated the effect of the detuning of the UV light on the loading efficiency of the optical dipole trap. Fig.~\ref{fig:NvsDetuning}~(a) shows the atom number yield at the end of the plain evaporation stage for three different magnetic field gradients of the compressed UV MOT. As the cloud thermalizes to the same temperature for all relevant detunings, the phase\hyp{}space density follows the same curve. The UV frequency detuning of $\delta\simeq\mathrm{-3.2}\Gamma_{\mathrm{3P}}$ for maximum atom number differs from the detuning for the maximum phase\hyp{}space density of the UV MOT, that was found to be $\delta\simeq\mathrm{-2.7}\Gamma_{\mathrm{3P}}$ \cite{Sebastian2014}. The residual differential ac Stark shift of the cooling transition induced by the $ 1070\, \textrm{nm} $ light cannot explain this difference, as it is small and increases the $\UVtransition $ transition frequency \cite{Duarte2011b, Safronova2012}. We attribute the observed difference to enhanced light\hyp{}assisted collisional loss on the UV transition, as the density is strongly increased in the presence of the optical dipole trap. In this case, the highest atom number is achieved for a larger laser detuning, which leads to a lower photon scattering rate and therefore a reduced atom loss from the trap.

%%% overlap duration   -----------
The duration of the combined cooling and optical dipole trapping phase affects the loading efficiency, as shown in Fig.~\ref{fig:NvsDetuning}~(b). Initially the atom number increases significantly with the overlap duration, demonstrating the efficiency of UV cooling in presence of the $1070\, \textrm{nm}$ trap light. After a few ms the atom number saturates and remains constant for about $10 \, \textrm{ms} $. We assume that during this time the influx of atoms from the UV MOT into the ODT is approximately balanced with two\hyp{}body losses occurring at high densities \cite{DePue2000}. 

Furthermore, we investigated the dependence of the transferred number of atoms on the optical power that is used to drive the UV MOT transition. A reduction from the typically used $12 \, \textrm{mW} $ for the $ 3 $ retro\hyp{}reflected beams shows that the experiment is not operated in the saturated regime. Remarkably, we find that even with just $ 1 \, \textrm{mW} $ of total UV power, an atom number of $2\times10^{6} $ is obtained in the optical trap before plain evaporation. For comparison we note that loading the ODT directly from the red compressed MOT, less than $1 \times 10^{5}$ atoms are transferred. 

In an effort to further improve the initial phase\hyp{}space density in the ODT, we utilized an additional trapping beam to enhance the confinement of the trap during the UV cooling stage. For this purpose the frequency of one of the ODT beams is shifted by $115\,\textrm{MHz}$ and directed back into the chamber to cross the existing trap under an angle of $51^{\circ}$. The additional beam has a maximum power of $ 15 \, \textrm{W} $ and a waist of $54 \, \textrm{$\mu$m}$ at the focus. This enhances the trap depth to around $ k_{B} \times 500 \, \textrm{$\mu$K} $ and the axial trap frequency to $ \omega_{z} = \left( 2\pi \right) \times 1.8 \, \textrm{kHz}$. However, with this configuration we find that the temperature along the axial direction of the atomic cloud before plain evaporation is increased by a factor of $3$, despite the continued UV cooling. Furthermore, after plain evaporation we found only a negligible enhancement of the phase\hyp{}space density of the atomic cloud. A possible reason is the interference between the ODT beams, as their polarizations are no longer mutually orthogonal and a spurious moving optical lattice is formed.

%%% loading description --------------

%%% Transport in a crossed ODT -------------------------------------------------------------------------------------
\section{Transport in a crossed ODT}

We realized a rapid long-distance transport of the atomic cloud over $ 25\, \textrm{cm} $ in $0.9\, \textrm{s}$ from the MOT chamber into the fused silica cell. This is implemented by moving the lens that simultaneously focuses and crosses the beams for the ODT by means of a compact air bearing translation stage (Nelson Air Corp., Atlas\hyp{}101\hyp{}310\hyp{}HD Stage), as illustrated in Fig.~\ref{fig:LithiumSetup}. For this configuration we ensure that the laser beams before the lens are aligned parallel to the direction of translation, so that the trapping geometry is maintained throughout the transport.

In order to realize the transport free of heating and excitation of center of mass oscillations, the trajectory is defined by a symmetric S\hyp{}curve velocity profile. The acceleration is linearly increased to $ 2.45\,\textrm{m}/\textrm{s}^2 $ in $ 0.45 \,\textrm{s}$ leading to a maximum velocity of $ 0.55 \,\textrm{m}/\textrm{s}$, directly followed by the deceleration. The large maximum acceleration in use is taking advantage of the strong axial confinement of the crossed optical dipole trap, as compared to single beam traps with comparable waist. The transport duration is only bound by technical limitations of the translation stage.

During the transport the optical power remains at the initial $ 22.5 \, \textrm{W} $ per beam. The measured heating rate for the trap kept stationary in the MOT chamber is $ 0.53(3) \, \textrm{$\mu$K} / \textrm{s} $, which is in good agreement with an estimate based on the spontaneous photon scattering rate $ \Gamma_{sc} $ at this optical power \cite{Grimm2000a}. As the observed heating rate is small compared to the initial temperature, a forced evaporation stage in the MOT chamber in order to reduce the optical power during the transport is therefore not required. We did not observe any significant additional heating or atom loss due to the transport, as compared to keeping the atomic cloud stationary. In order to measure low heating rates, the atomic cloud was first evaporatively cooled, and the temperature was measured after multiple repetitions of the transport trip in a single run. From this measurement a small additional heating rate of approximately $ 0.04 \, \textrm{$\mu$K / \textrm{trip}} $ induced by a single transport at full optical power was determined.

Table~\ref{tab:TransportTable} shows the parameters of our crossed beam transport configuration in comparison with a transport based on a single beam optical dipole trap. If equal laser power and axial confinement strength are assumed the crossed beam trap offers the advantage of larger beam diameters for an enhanced loading efficiency \cite{Burchianti2014}, and an improved aspect ratio. With the availability of efficient cooling in the UV MOT the larger volume of the crossed beam configuration is favored over the larger trap depth of the single beam trap for the loading process. The larger photon scattering rate of the single beam trap can be avoided by using less power. However, to maintain the same axial confinement and to limit the aspect ratio, an even smaller beam waist is required in this case. Using such a low trapping volume configuration then requires additional measures to obtain a good loading efficiency \cite{Zimmermann2011}.

%%% Evaporation to degeneracy -------------------------------------------------------------------------------------
\section{Evaporation to degeneracy}

In the following we describe how quantum degeneracy in the trapped lithium sample is achieved in the same crossed beam configuration as previously used for capturing atoms from the UV MOT and for optical transport. For this trap we give examples of optimized evaporative cooling schemes for the cases of weak and strong interactions in the Fermi gas. 

To produce a weakly interacting quantum degenerate Fermi gas, we apply a magnetic bias field of $ 330 \, \textrm{G} $ to increase the $ \left|1  \right\rangle \textrm{-} \left|2  \right\rangle$  s\hyp{}wave scattering length to $ a_{12} = -290 \, a_{0} $ \cite{Zurn2013}. To reduce the trap depth $U_0$ for forced evaporation, we reduce the optical power by means of AOMs, which are introduced into each of the two beams. Logarithmic photodetectors and high bandwidth PID regulators allow the stabilization and control of the optical power over four orders of magnitude. 

The time\hyp{}dependent trap depth during evaporation follows the theoretical curve $ U_0\left(t\right) = U_0 \left( 0\right) \left( 1+ t/\tau \right)^{-2\left(\eta^{\prime}-3\right) / \eta^{\prime}} $ resulting in a constant truncation parameter $ \eta$ throughout the evaporation process \cite{OHara2001}. We use a time constant of $\tau = 4.2\,s$, which is experimentally optimized for maximum atom number, and $\eta^{\prime} = 10.8$ corresponding to a truncation parameter of $\eta = 10$.

\begin{figure}
\includegraphics[width=8.0cm]{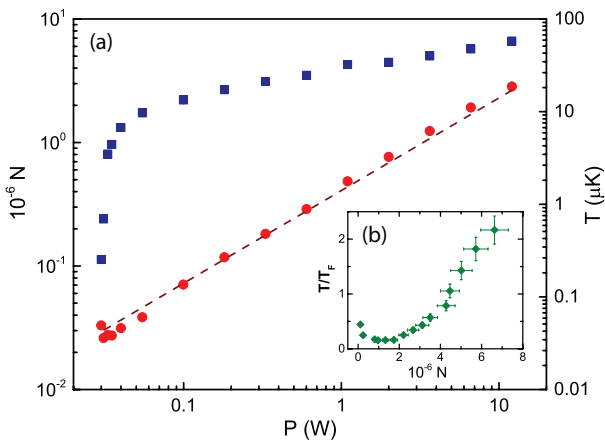}%
\caption{(Color online) (a) Total number of atoms $N$ (blue squares) and temperature $T$ (red filled circles) at various points during forced evaporative cooling performed at $330 \, \textrm{G}$. $ P $ is the optical power per beam of the ODT at which the evaporation ramp was halted. The dashed line corresponds to $0.1\,U_{0} / k_{B}$, which is in close agreement with the measured temperature of the atomic cloud, indicating an approximately constant truncation $ \eta $ parameter throughout the evaporation. (b) $T/T_{\text{F}}$ vs $N$ for the same measurement. The error shown in (b) includes, in addition to the statistical error of $10$ repetitions for each point, a systematic uncertainty of $ 10 \, \% $ in the measured quantities.}
\label{fig:NTvsP}
\end{figure}

To characterize the evaporation we probe the atomic cloud at various intermediate optical powers during the ramp. For this purpose, the magnetic field and the optical potential are rapidly switched off and the density distribution is determined by standard time\hyp{}of\hyp{}flight (TOF) absorption imaging. The projected density profiles are fitted to a Fermi distribution function, from which we extract atom number, temperature, and $ T / T_F $. The results are shown in Fig.~\ref{fig:NTvsP}. We observe efficient cooling of the cloud down to a laser power of $P=50 \, \textrm{mW} / \textrm{beam} $, where $8.7\times10^{5}$ atoms in each spin state remain trapped at a temperature of $60 \, \textrm{nK}$, corresponding to $T/T_{\text{F}}=0.16\left(2 \right)$. 

To create a strongly interacting quantum gas in the BEC\hyp{}BCS crossover, evaporation is performed near the broad Feshbach resonance. Specifically, for the production of a molecular BEC we generally follow the procedure described by Jochim \textit{et al.} \cite{Jochim2003}. The magnetic field is ramped to $ 770 \, \textrm{G} $, where the s\hyp{}wave scattering length reaches approximately $ a_{12} = 5100 \, a_{0} $, and $k_{F}a_{12} = 4.8$, where $k_{F}$ is the Fermi wave vector. The binding energy of the Feshbach molecules at this magnetic field is on the order of $ k_{B} \times 1.1 \, \textrm{$\mu$K} $ \cite{Zurn2013}.  

The strongly interacting atomic cloud is evaporatively cooled by reducing the trap depth according to $ U_0 \left(t\right) = U_0 \left( 0\right) \left( 1 - t/\tau_u \right)^{2\left(\eta^{\prime}-3\right) / \left( \eta^{\prime} - 6 \right) } $, with a time constant $ \tau_u = 3 \, \textrm{s} $. This evaporation trajectory maintains a constant truncation parameter and was derived by Luo $\etal$ \cite{Luo2006} for a Fermi gas mixture with a unitarity\hyp{}limited s\hyp{}wave scattering cross section. During the evaporation, weakly bound dimers are formed by three\hyp{}body recombination leading to an almost pure sample of Feshbach molecules for temperatures well below their binding energy \cite{Jochim2003, Petrov2003b, Chin2004a}.

\begin{figure}
\includegraphics[width=8.0cm]{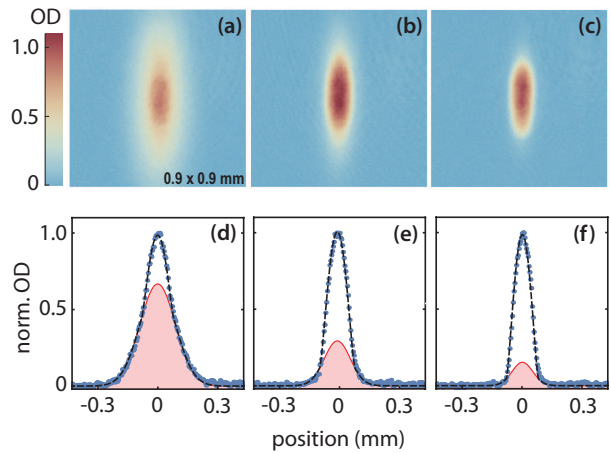}%
\caption{(Color online) Density distribution and individually normalized horizontal central profiles of the $ {}^{6}\textrm{Li}_2 $ cloud. Forced evaporative cooling is performed at $ 770 \, \textrm{G} $ to a final optical power of (a,d) $ 100\, \textrm{mW}$, (b,e) $ 45\, \textrm{mW} $, and (c,f) $ 35\, \textrm{mW} $ per beam of the ODT. The optical density (OD) of the cloud is recorded after a TOF of $ 6 \, \textrm{ms} $ at a magnetic field of $ 690 \, \textrm{G} $. Each central profile (blue filled circles) is normalized to one and the bimodal fitting function (dashed line) consists of a parabolic curve added to a Gaussian. The Gaussian component (solid line) illustrates the bimodal character of the density distribution.}
\label{fig:BECtransition}
\end{figure}

To increase the visibility of the bimodal density distribution, which is characteristic for the onset of Bose\hyp{}Einstein condensation, the interaction strength between the molecules is lowered by linearly reducing the magnetic field to $ 690 \, \textrm{G}$ in $ 100\, \textrm{ms} $ before TOF absorption imaging \cite{Burchianti2014, Gunton2013a}. This reduces the atomic scattering length $a_{12}$ to around $1400 \, a_0 $, which is related to the molecular s\hyp{}wave scattering length via $ a_{\mathrm{mol}} = 0.6 \, a_{12} $  \cite{Petrov2004}. At this magnetic field absorption imaging of the molecules with light on the atomic imaging transition is still efficient \cite{Bartenstein2004}. The trap depth $U_{0}^{\mathrm{mol}}$ for the molecules is assumed to be twice as large as compared to the case for the atoms \cite{Jochim2003}. At $U_{0}^{\mathrm{mol}} / k_{B} = 2.9 \, \textrm{$\mu$K} $ we observe  $ 1.1 \times 10^6 $ molecules and a deviation from a thermal density distribution indicating the onset of condensation (see Fig.~\ref{fig:BECtransition}). At a final trap depth of $U_{0}^{\mathrm{mol}} / k_{B} = 1.0 \, \textrm{$\mu$K} $ we observe BEC containing around $ 3 \times 10^5 $ Feshbach molecules, which is inferred from a fit to the bimodal density distribution. At this point the trap frequencies are $\omega_{z} = \left(2 \pi \right) \times 7.8 \, \textrm{Hz} $ and $\bar{\omega}_r = \left(2 \pi \right) \times 120 \, \textrm{Hz} $, leading to a critical temperature of $ T_c = 170 \, \textrm{nK} $ for an ideal Bose gas \cite{Jochim2003}. The axial confinement is enhanced by the residual curvature of the applied magnetic bias field, leading to a measured contribution to the axial trap frequency of $ \left(2 \pi \right) \times 4.6 \, \textrm{Hz} $. The effect of the magnetic field curvature on the radial trap frequency is neglected.

%%% Conclusion -------------------------------------------------------------------------------------
\section{Conclusion}
In conclusion we have demonstrated a refined scheme for an all\hyp{}optical production of a quantum degenerate $\Li$ gas that involves long\hyp{}distance transport for good optical access in a glass cell. Here we are taking full advantage of the high phase-space density obtained by the UV MOT for lithium and the possibility of continued laser cooling in the optical dipole trap. This allows an improved loading efficiency into optical dipole traps with lower trap depth and hence comparatively large volume. The benefit of UV cooling is demonstrated to be present even for low optical power. In addition to loading with high atom number, our crossed beam configuration also offers sufficient axial confinement for a rapid and near\hyp{}adiabatic transport of the atomic cloud. By means of the same trap, evaporative cooling is performed to obtain a large quantum degenerate Fermi gas and a molecular BEC, which shows the efficiency of the approach. Our implementation facilitates experiments requiring unconstrained optical access, such as quantum gas microscopy. In combination with the available UV laser, the fast production of weakly bound dimers is a good starting point for experimental investigation of long-range molecular states near the $2S\textrm{-}3P$ dissociation limit. 

%%% Acknowledments -------------------------------------------------------------------------------------
This research is supported by the National Research Foundation, Prime Ministers Office, Singapore and the Ministry of Education, Singapore under the Research Centres of Excellence program.

\bibliographystyle{unsrt}

\end{document}